# Dynamic strain-mediated coupling of a single diamond spin to a mechanical resonator

Preeti Ovartchaiyapong†

Department of Physics, University of California Santa Barbara, Santa Barbara, CA 93106

Kenneth W. Lee†

Department of Physics, University of California Santa Barbara, Santa Barbara, CA 93106

Bryan A. Myers

Department of Physics, University of California Santa Barbara, Santa Barbara, CA 93106

Ania C. Bleszynski Jayich*

Department of Physics, University of California Santa Barbara, Santa Barbara, CA 93106

Email: ania@physics.ucsb.edu

†These authors contributed equally to this work

*Corresponding author

# Abstract

The development of hybrid quantum systems is central to the advancement of emerging quantum technologies, including quantum information science and quantum-assisted sensing. The recent demonstration of high quality single-crystal diamond resonators has led to significant interest in a hybrid system consisting of nitrogen-vacancy center spins that interact with the resonant phonon modes of a macroscopic mechanical resonator through crystal strain. However, the nitrogen-vacancy spin-strain interaction has not been well characterized. Here, we demonstrate dynamic, strain-mediated coupling of the mechanical motion of a diamond cantilever to the spin of an embedded nitrogen-vacancy center. Via quantum control of the spin, we quantitatively characterize the axial and transverse strain sensitivities of the nitrogen-vacancy ground state spin. The nitrogen-vacancy center is an atomic scale sensor and we demonstrate spin-based strain imaging with a strain sensitivity of $3 \times 10^{-6}$ strain $\cdot$ Hz$^{-1/2}$. Finally, we show how this spin-resonator system could enable coherent spin-phonon interactions in the quantum regime.

# Introduction

Nitrogen-vacancy (NV) centers in diamond are amongst the most promising implementations of quantum bits for quantum information processing[1] and nanoscale field sensors[2]. Because of their excellent coherence properties[3] and their ability to coherently couple to various external fields, NV centers are ideal candidates for integration into a hybrid quantum system, which combines the merits of two or more physical systems to mitigate their individual weaknesses. To date, experimental efforts have focused on integrating the NV center into hybrid systems consisting of photons[4,5,6], nuclear spins[7], and magnetically coupled mechanical resonators[8,9].

Recent demonstrations of high quality (Q) factor single-crystal diamond (SCD) mechanical resonators[10,11] have generated interest in the development of a monolithic structure consisting of such a resonator that is strain-coupled to the electronic spin of an embedded NV center[12,13]. With the NV's long quantum coherence time and diamond's low mechanical losses, this system provides a unique environment to engineer spin-spin entanglement or spin squeezing[12]. Moreover, a coherent spin-phonon interaction may be used to generate non-classical states of the resonator, such as Schrödinger cat states[13,14,15,16]. An inherent advantage of a strain coupled spin-resonator system is the intrinsic nature of the coupling; no additional components are required to tune the coupling strength, such as an atomic force microscope tip or cavity, which may introduce fluctuating fields that decohere the NV spin. Additionally, in contrast with other hybrid mechanical systems, this NV-resonator system requires no functionalization of the resonator with, *e.g.* a mirror or electrode, which could adversely affect the Q.

However, to engineer such a system, we require a quantitative understanding of how strain couples to the ground state spin levels of the NV center. Despite the NV's well understood sensitivity to magnetic[2] and electric[17,18] fields and temperature[19,20], the NV ground state strain sensitivity has not been fully characterized, and there exist competing theoretical models of the strain interaction[18,21,22,23].

In this letter, we use high-Q SCD cantilevers to probe the interaction between a single NV spin and crystal strain. The magnitude of the strain is highly controlled, and is set by the cantilever's mode shape and amplitude of motion, the depth and lateral location of the NV in the cantilever, and the orientation of the NV with respect to the cantilever axis. The strain induced by an oscillating cantilever modulates the energy of the ground state spin levels, which is detected via coherent quantum control of the spin[24]. Combined with the known strain profile of a cantilever under pure bending and the four possible NV orientations in the diamond crystal structure, our measurements allow us to extract the axial and transverse strain susceptibilities. Thus the NV is a calibrated strain sensor, with which we image the strain profile of the resonator. Importantly, we demonstrate NV-strain coupling in a geometry appropriate for realizing a hybrid quantum system, and show that by scaling down the dimensions of the resonator, it should be possible to reach the quantum regime.

## Results

### Strain model and experimental setup

Here, we model strain by an effective electric field in which strain induced displacements of the atoms alter the electron density in the crystal, resulting in local electric fields[17,18]. Therefore, the ground state Hamiltonian for an NV in the presence of magnetic and strain fields may be described by ($h = 1$)

$$H_{\text{NV}} = (D_0 + d_\| \epsilon_\|)S_z^2 + \gamma_{\text{NV}}\mathbf{S} \cdot \mathbf{B} - \frac{d_\perp \epsilon_\perp}{2}\left(e^{-i\phi_s}S_+^2 + e^{i\phi_s}S_-^2\right) \quad (1)$$

where $D_0$ is the crystal-field splitting ($D_0 \approx 2.87$ GHz), $\gamma_{\text{NV}}$ is the gyromagnetic ratio ($\gamma_{\text{NV}} \approx 2.8$ MHz·G$^{-1}$), $d_\|$ and $d_\perp$ are the strain susceptibility parameters parallel and perpendicular to the NV symmetry axis (whose values are determined in this work), $S_\pm$ are the spin-1 raising and lowering operators, $\epsilon_\| = \epsilon_z$, $\epsilon_\perp = \sqrt{\epsilon_x^2 + \epsilon_y^2}$, $\tan(\phi_s) = \epsilon_y/\epsilon_x$, and $\{\epsilon_i\}_{i=x,y,z}$ are the diagonal components of the strain tensor defined in the NV's basis (note we have neglected shear). We will henceforth refer to $d_\|$ and $d_\perp$ as the axial and transverse strain susceptibility parameters. A uniaxial strain along the NV symmetry axis, $\epsilon_\|$, results in a shift of the crystal-field splitting, whereas a strain transverse to the symmetry axis, $\epsilon_\perp$, results in mixing between the $|m_s = \pm 1\rangle$ $S_z$ eigenstates. Because strain and stress are tensors, the uniaxial stress due to the bending of the cantilever will generate strain along all three principal directions due to the Poisson effect[25]. This effect cannot be neglected, as shown later.

In this experiment, an NV interacts with the fundamental mechanical mode of the cantilever via strain (Fig. 1a). For small displacements of the cantilever, the strain felt by the NV center is linear in cantilever position, and the total strain can be written as $\epsilon = \epsilon_0 \frac{X_d}{x_0}$, where $X_d$ is the amplitude of driven motion, $x_0$ is the amplitude of zero point motion and $\epsilon_0$ is the strain induced by the zero point motion[12]. If the NV is in the presence of a DC magnetic field aligned closely to its symmetry axis, equation (1) can be approximately diagonalized with eigenstates $\{|0\rangle, |\pm\rangle\}$, and the energies can be expressed as a function of the normalized beam displacement, $X = \frac{X_d}{x_0}$

$$E_0 \simeq 0 \; ; \; E_\pm(X) \simeq D_0 + g_\parallel X \pm \sqrt{(\gamma_{NV} B_z)^2 + (g_\perp X)^2} \tag{2}$$

where $g_\parallel = d_\parallel \epsilon_{0\parallel}$ and $g_\perp = d_\perp \epsilon_{0\perp}$ are the axial and transverse strain couplings to the zero point motion of the resonator, and $|\pm\rangle$ are defined as (Supplementary Note 1),

$$|0\rangle = |0\rangle \; ; \; |\pm\rangle = \cos\frac{\alpha}{2}|1\rangle \mp e^{-i\phi_s}\sin\frac{\alpha}{2}|-1\rangle \tag{3}$$

where $\alpha/2$ is the Stückelberg angle and $\tan(\alpha) = d_\perp \epsilon_\perp / \gamma_{NV} B_z$. We note that by tuning $B_z$ we may selectively characterize axial and transverse strain, as evidenced in equation (2).

The experiment consists of a single spin embedded in a SCD cantilever (Fig. 1b). The cantilevers used here (Fig. 1c) were fabricated using the technique described in ref. 10, and are 60 μm long, 15 μm wide, and ~ 1 μm thick. They have fundamental frequencies $\omega_m/2\pi$ near 1 MHz and quality factors exceeding $3.0 \times 10^5$, which were measured using optical interferometry[10]. All measurements were performed at room temperature and at pressures of approximately $10^{-5}$ torr. NVs were formed approximately 50 nm below the diamond surface via

$^{14}$N ion implantation and vacuum annealing (see Methods). The orientation of the NV with respect to the cantilever (Supplementary Fig. 1) is identified through its Zeeman splitting in an applied DC magnetic field. We align this field closely to the NV axis to break the degeneracy of the $|\pm 1\rangle$ spin levels. We encode our qubit in the $|0\rangle$ and $|+\rangle$ spin states, and coherently manipulate the spin using near-resonant microwaves. We are able to distinguish the effects of $\epsilon_\parallel$ and $\epsilon_\perp$ by noting an AC $\varepsilon_\parallel$ modulates the qubit frequency at $\omega_m$, whereas an AC $\varepsilon_\perp$ modulates the qubit frequency at approximately $2\omega_m$.

**Axial strain detection**

Figure 2a shows the measurement protocol for axial strain detection. The cantilever is resonantly driven with a piezoelectric transducer, and the phase of the drive is not phase-locked to the timing of the control pulses. The NV spin is initialized to the $|0\rangle$ state via optical pumping with a 532 nm laser. Microwaves tuned to the $|0\rangle \leftrightarrow |+\rangle$ transition are applied to the NV to carry out a Hahn echo pulse sequence with a total free evolution time $T = 2\tau$. During the free evolution periods, the resonantly driven motion of the cantilever imprints a relative phase onto the qubit, which is then converted into a population difference and read out via the spin-dependent fluorescence. In the measurements of axial strain, $B_z$ was set to 22 G to suppress the effects of transverse strain, as shown in equation (2).

An AC axial strain results in a modulation of the spin population difference that is correlated to the motion of the cantilever. As depicted in Fig. 2b, when the total evolution time of the spin is equal to an odd integer multiple of the cantilever period, the accumulated phase is maximized, resulting in a reduced population in $|0\rangle$. When the total evolution time is equal to an even multiple of the cantilever period, the qubit acquires no phase, resulting in a revival of $|0\rangle$ state population. Because the cantilever's motion and microwave pulse timing are not phase locked, the signal averages over a uniformly distributed initial phase of the motion, and is given by a zero order Bessel function (Supplementary Note 2).

By fitting the data in Fig. 2b (solid black line) to the expected echo signal, we extract an axial strain coupling $G_\parallel = g_\parallel X = 2.1 \pm 0.1$ MHz and a cantilever frequency $\omega_m/2\pi = 882.0 \pm 2.0$ kHz, in agreement with the driving frequency of the cantilever of 884.583 kHz. The strain coupling can be tuned with cantilever drive. Fig. 2c shows a Hahn echo measurement of a single NV for different beam deflection amplitudes. In the absence of cantilever drive, the spin retains its coherence over a single period of the cantilever's motion. As the amplitude of driven motion increases, the increased strain coupling results in a reduced $|0\rangle$ state population. For sufficiently high couplings, Bessel fringe oscillations are observed, indicating that the spin has precessed more than once around the equator of the Bloch sphere. In Fig. 2d, we plot $G_\parallel$ for four different drive amplitudes and confirm that the coupling is linear with beam displacement.

**Single spin strain imaging**

With its atom-sized spatial extent, the NV is a novel, nanoscale strain sensor. We demonstrate NV-based strain imaging by measuring $G_\parallel$ for several NVs located at different positions along the length of the cantilever (Supplementary Fig. 2). In Fig. 3a, we show the simulated strain profile of a cantilever under pure bending. Fig. 3b shows the measured strain couplings for several NVs along the cantilever for a fixed drive. The data are in good agreement with the theoretical strain profile, shown by the shaded region in Fig. 3b (Supplementary Note 3). Notably, this agreement provides convincing evidence that strain is responsible for the spin evolution. From the data in Fig. 3b, we can extract $d_\parallel$ by combining the interferometrically measured amplitude of motion, the NV's position and orientation in the cantilever, and the expected strain profile of the cantilever (Supplementary Note 4). By measuring several NVs, we average over the variations in NV depth and local strain inhomogeneities, and we find an average value of $d_\parallel = 13.4 \pm 0.8$ GHz $\cdot$ strain$^{-1}$, and hence calibrate our strain sensor. Given our current experimental parameters, we estimate an axial strain sensitivity of approximately $3 \times 10^{-6}$ strain $\cdot$ Hz$^{-1/2}$ (Supplementary Note 5). For the cantilevers in this experiment, an NV at the base could then detect $\sim 7$ nm of motion in a 1 second measurement.

**Transverse strain detection**

We next measured the transverse strain coupling. To do so, we measured NVs oriented perpendicular to the cantilever stress, as shown in Fig. 4a. Although the stress is entirely perpendicular to NVs oriented $[\bar{1}1\bar{1}]$ and $[1\bar{1}\bar{1}]$ (shown with red bonds), these NVs will experience non-negligible axial strain in addition to transverse strain due to the Poisson effect[25]. Experimentally, the effects of axial and transverse strains are identified by beatnotes present in the spin evolution. To enhance our sensitivity to these beatnotes, we applied an XY-4 control sequence[26] (Fig. 4b) to the NV and decreased $B_z$ to 16 G. This sequence increases our sensitivity by extending the coherence time, correcting for 1$^{st}$ order timing errors, and increasing the interrogation time. The data and fit to the expected XY-4 signal (Supplementary Note 6) in Fig. 4c give $G_\perp = 6.7 \pm 0.4$ MHz, $G_\parallel = (2.7 \pm 0.5) \times 10^2$ kHz and a cantilever frequency $\omega_m/2\pi = 899 \pm 8$ kHz, in agreement with the cantilever drive frequency of 884.890 kHz. Transforming the cantilever strain tensor into the NV's coordinate system and using the Poisson ratio of .069 for CVD diamond[24] (Supplementary Note 4), we extract $d_\parallel = 13.3 \pm 1.1$ GHz $\cdot$ strain$^{-1}$ and $d_\perp = 21.5 \pm 1.2$ GHz $\cdot$ strain$^{-1}$. The extracted value of $d_\parallel$ is in very good agreement with the extracted value of $d_\parallel$ measured for the NVs experiencing predominantly axial strain in Fig. 3.

## Discussion

The measured strain susceptibility parameters indicate that while the simple model equating strain and electric field accurately predicts the qualitative nature of the NV-strain coupling, it fails to capture the quantitative nature of the strain susceptibility parameters. In particular, this model predicts a value of $d_\parallel$ that is approximately 50 times smaller than $d_\perp$[18], while our measurements show that $d_\parallel$ is approximately 1.5 times smaller than $d_\perp$. During the preparation of this manuscript, a closely related study[27] reported measurements of $d_\parallel = 5.46 \pm 0.31 \text{ GHz} \cdot \text{strain}^{-1}$ and $d_\perp = 19.63 \pm 0.40 \text{ GHz} \cdot \text{strain}^{-1}$ using static bending of diamond cantilevers. In comparison with our results, we find good agreement with their value of $d_\perp$ and find that their value of $d_\parallel$ is approximately 3 times smaller than our measured value. We make several observations in regards to this discrepancy. First, ref. 27 reports large variations (up to one order of magnitude) in the strain susceptibility parameters across NVs. Second, the AC sensing techniques used in our work are inherently more sensitive than the techniques used in ref. 27, and hence the amplitude of cantilever bending is significantly smaller here. Large bending amplitudes could lead to deviations from Euler-Bernoulli theory. Furthermore the resonance phenomenon we utilize provides a well-defined mode shape, whereas DC bending may not isolate a single mode. Another study[23] measured shifts of the NV ground state crystal field splitting under hydrostatic pressure, and their measured value of $d_\parallel = 17.5 \text{ GHz} \cdot \text{strain}^{-1}$ is in good agreement with our result. However, no value of $d_\perp$ was reported. *Ab initio* calculations from ref. 23 estimate a value of $d_\parallel = 7.44 \text{ GHz} \cdot \text{strain}^{-1}$ by considering modifications of the spin-spin coupling due to changes in the NV bond lengths. This value is of the same order of magnitude as our result, indicating that this effect plays a significant role in the NV-strain interaction.

Looking toward quantum applications[12], our cantilever's current zero point motion coupling of $g_\perp \sim 0.04$ Hz may be significantly increased by miniaturizing the resonator. For instance, a doubly-clamped beam with dimensions of (2 μm, 50 nm, 50 nm) and $\omega_m/2\pi =$ 238 MHz has a coupling $g_\perp \approx 180$ Hz. While having smaller strain couplings than, for example, quantum dot systems where the coupling can be hundreds of kHz[28], the NV center benefits from its long $T_2$ coherence times. Assuming $T = 100$ mK, a mechanical $Q = 10^6$, and a $T_2 = 100$ ms, this doubly clamped beam achieves a single spin cooperativity, $\eta = 2\pi \frac{g_\perp^2 T_2}{\gamma \bar{n}} \approx 1.6$, where $\gamma = \omega_m/Q$ is the damping rate and $\bar{n}$ is the average thermal occupation number of the mechanical mode. We note that although smaller resonators would produce a larger $g_\perp$, clamping losses would limit their $Q$s[29]. At $T \lesssim 50$ K, SCD resonators have demonstrated[2] $Q > 10^6$. In the devices described here, the Hahn echo $T_2$ times are roughly 10 μs. Using NVs formed by a nitrogen delta-doping technique[30] and applying higher order dynamical decoupling, longer $T_2$ times are possible[31] and recently, a $T_2 > 500$ ms has been measured at low temperatures[32]. A cooperativity $\eta > 1$ enables experiments in the quantum regime that are mediated by a coherent spin-phonon interaction[12]. We note that the strain coupling to the orbital levels of the excited state is significantly higher[4,13], and for the resonator described above, would give a coupling $g_\perp$ of several MHz, which presents an alternative route to accessing the quantum regime.

In conclusion, we have presented a novel hybrid system in which a single NV center spin is strain-coupled to the mechanical motion of a macroscopic cantilever. We quantitatively measured the NV ground state spin coupling to axial and transverse strain, and these results will guide theoretical investigations of the strain coupling mechanism. While our measurements indicate that shear does not couple significantly to the ground state spin levels of the NV center, we note that the cantilever-based technique presented here can be extended in future experiments to probe effects of shear on the NV. Furthermore, the coherent strain detection approach we have presented can be extended to other solid state spin systems, such as SiC color centers[33,34]. With its atomic scale spatial resolution, NV-based strain sensing could be useful for identifying the poorly understood origins of dissipation in mechanical resonators, such as anchoring losses, which will be instrumental in engineering higher Q resonators[35]. Moreover, our results will provide insight into the role of strain-induced decoherence that could ultimately limit the effectiveness of many NV-based applications[12,36,37]. Most importantly, we demonstrated that our system provides a feasible route to accessing coherent spin-phonon interactions, with applications ranging from quantum control of a macroscopic resonator to long-range spin-spin interactions.

## Methods

**Experimental setup and sample preparation**

The experimental setup consists of a home-built confocal microscope with an integrated vacuum chamber. The experiments are done at room temperature and high vacuum ($\approx 10^{-5}$ torr) to reduce mechanical dissipation in the resonators. A continuous wave laser at 532 nm is used for optical pumping and readout of the NV spin, and is gated with an AOM. Photons emitted by the NV into the phonon sideband are collected into a single-mode fiber and directed to a fiber-coupled avalanche photodiode. For strain measurements, the degeneracy of the $m_s = \pm 1$ spin levels is broken by a DC magnetic field supplied by a rare earth magnet. Microwaves used for resonant spin manipulation are delivered to the sample by a wire that sits approximately 30 μm above the sample. Phase control of the microwave pulses is accomplished via IQ modulation. Timing of the pulse sequences is controlled by a Spincore Pulseblaster ESR-Pro 500 MHz card.

The cantilevers used in this experiment were fabricated using the techniques described in ref. 10. NVs were formed by $^{14}$N ion implantation with a dosage of $3 \times 10^9$ ion·cm$^{-2}$ at 40 keV and a 0° tilt, which yields an expected depth of 51.5 nm (calculated by SRIM). The sample was annealed under high vacuum (~$10^{-6}$ torr) at 800 °C for 3 hours.

The cantilevers are resonantly driven with a piezoelectric transducer that is clamped to the sample holder and electrically driven with a lock-in amplifier (Zurich Instruments). The cantilevers' mechanical properties are characterized using standard optical interferometry techniques, as described in ref. 10. For our experiments, it is very important that we know the amplitude of driven motion of the cantilever. This is calibrated by finding the piezo voltage at which the tip of the cantilever hits the silicon substrate that sits 1 μm below the cantilever. This is measured by an interruption of the Bessel fringes in the interferometric signal.

**Error Analysis**

Errors in the extracted axial strain coupling parameters and resonator frequencies are given by the standard error in the fit where the couplings and the resonator frequency are the only free parameters. The error in the quoted axial strain susceptibility parameter is given by the standard error in the average value of those parameters as measured by nine NVs. The error in the transverse strain coupling and susceptibility parameter are given by the measurement of a single NV, and the standard error of the fit shown in Fig. 4c.

**Acknowledgements**

The authors wish to thank Shimon Kolkowitz, Gino Graziano, Laetitia Pascal, Steven Bennett and Adam Gali for helpful discussions. This work is supported by the Air Force Office of Scientific Research MURI program, DARPA QuASAR and the UCSB MRSEC. B.A.M. is supported through a fellowship from the Department of Defense (NDSEG).

**Author Contributions**

P. O. designed and fabricated the device. K.W.L. and P.O carried out the experiments.

K.W.L. and B.A.M. contributed to the software infrastructure. K.W.L. and B.A.M. analyzed the data and P.O. provided feedback. K.W.L and A.J wrote the paper with feedback from all authors. A.J. supervised the project. All authors contributed to the design of the experiment and discussions during the course of the measurements and analysis.

**Competing financial interests**

The authors declare no competing financial interests.

# References

1. Childress, L. & Hanson, R. Diamond NV centers for quantum computing and quantum networks. *MRS Bull.* **38,** 134–138 (2013).

2. Rondin, L. *et al.* Magnetometry with nitrogen-vacancy defects in diamond. *Rep. Prog. Phys.* **77,** 056503 (2014).

3. Balasubramanian, G. *et al.* Nanoscale imaging magnetometry with diamond spins under ambient conditions. *Nature* **455,** 648–51 (2008).

4. Togan, E. *et al.* Quantum entanglement between an optical photon and a solid-state spin qubit. *Nature* **466,** 730–4 (2010).

5. Bernien, H. *et al*. Heralded entanglement between solid-state qubits separated by three metres. *Nature* **497**, 86–90 (2013).

6. Faraon, A., Barclay, P. E., Santori, C., Fu, K. C. & Beausoleil, R. G. Resonant enhancement of the zero-phonon emission from a colour centre in a diamond cavity. *Nature Photonics* **5**, 301–305 (2011).

7. Childress, L. *et al*. Coherent dynamics of coupled electron and nuclear spin qubits in diamond. *Science* **314**, 281–5 (2006).

8. Kolkowitz, S. *et al*. Coherent sensing of a mechanical resonator with a single-spin qubit. *Science* **335**, 1603–1606 (2012).

9. Rabl, P. *et al*. Strong magnetic coupling between an electronic spin qubit and a mechanical oscillator. *Phys. Rev. B* **79**, 041302 (2009).

10. Ovartchaiyapong, P., Pascal, L. M. A., Myers, B. A., Lauria, P. & Bleszynski Jayich, A. C. High quality factor single-crystal diamond mechanical resonators. *Appl. Phys. Lett.* **101,** 163505 (2012).

11. Tao, Y., Boss, J. M., Moores, B. a & Degen, C. L. Single-crystal diamond nanomechanical resonators with quality factors exceeding one million. *Nat. Commun.* **5,** 3638 (2014).

12. Bennett, S.D. *et al*. Phonon-Induced Spin-Spin Interactions in Diamond Nanostructures: Application to Spin Squeezing. *Phys. Rev. Lett.* **110**, 156402 (2013).

13. Kepesidis, K. V., Bennett, S. D., Portolan, S., Lukin, M. D. & Rabl, P. Phonon cooling and lasing with nitrogen-vacancy centers in diamond. *Phys. Rev. B* **88,** 064105 (2013).

14. Wilson-Rae, I., Zoller, P. & Imamoğlu, A. Laser cooling of a nanomechanical resonator mode to its quantum ground state. *Phys. Rev. Lett.* **92**, 075507 (2004).

15. O'Connell, A. D. *et al*. Quantum ground state and single-phonon control of a mechanical resonator. *Nature* **464,** 697–703 (2010).

16. Teufel, J. D. *et al.* Sideband cooling of micromechanical motion to the quantum ground state. *Nature* **475,** 359–63 (2011).

17. Van Oort, E. & Glasbeek, M. Electric-field-induced modulation of spin echoes of N-V centers in diamond. *Chem. Phys. Lett.* **168,** 529–532 (1990).

18. Dolde, F. *et al.* Electric-field sensing using single diamond spins. *Nature Physics* **7,** 459–463 (2011).

19. Acosta, V. M. *et al.* Temperature Dependence of the Nitrogen-Vacancy Magnetic Resonance in Diamond. *Phys. Rev. Lett.* **104,** 070801 (2010).

20. Toyli, D. M., de las Casas, C. F., Christle, D. J., Dobrovitski, V. V & Awschalom, D. D. Fluorescence thermometry enhanced by the quantum coherence of single spins in diamond. *Proc. Natl. Acad. Sci. U. S. A.* **110**, 8417–21 (2013).

21. Maze, J. R. *et al.* Properties of nitrogen-vacancy centers in diamond: the group theoretic approach. *New J. Phys.* **13,** 025025 (2011).

22. MacQuarrie, E. R., Gosavi, T. A., Jungwirth, N. R., Bhave, S. A. & Fuchs, G. D. Mechanical Spin Control of Nitrogen-Vacancy Centers in Diamond. *Phys. Rev. Lett.* **111**, 227602 (2013).

23. Doherty, M. W. *et al.* Electronic Properties and Metrology Applications of the Diamond NV-Center under Pressure. *Phys. Rev. Lett.* **112,** 047601 (2014).

24. de Lange, G., Wang, Z. H., Ristè, D., Dobrovitski, V. V & Hanson, R. Universal dynamical decoupling of a single solid-state spin from a spin bath. *Science* **330**, 60–3 (2010).

25. Klein, C. A. & Cardinale, G. F. Young's modulus and Poisson's ratio of CVD diamond. *Diam. Relat. Mater.* **2,** 918–923 (1993).

26. de Lange, G., Ristè, D., Dobrovitski, V. V. & Hanson, R. Single-Spin Magnetometry with Multipulse Sensing Sequences. *Phys. Rev. Lett.* **106**, 080802 (2011).

27. Teissier, J., Barfuss, A., Appel, P., Neu, E. & Maletinsky, P. Resolved sidebands in a strain-coupled hybrid spin-oscillator system. Preprint at http:// http://arxiv.org/abs/1403.3405 (2014).

28. Yeo, I. *et al.* Strain-mediated coupling in a quantum dot – mechanical oscillator hybrid system. *Nature Nanotechnology* **9,** 106–110 (2014).

29. Cole, G. D., Wilson-Rae, I., Werbach, K., Vanner, M. R. & Aspelmeyer, M. Phonon-tunnelling dissipation in mechanical resonators. *Nature Communications* **2,** 231 (2011).

30. Ohno, K. *et al.* Engineering shallow spins in diamond with nitrogen delta-doping. *Appl. Phys. Lett.* **101,** 082413 (2012).

31. Myers, B. A. *et al.* Probing surface noise with depth-calibrated spins in diamond. Preprint at http://arxiv.org/abs/1402.5392 (2014).

32. Bar-Gill, N., Pham, L. M., Jarmola, A, Budker, D. & Walsworth, R. L. Solid-state electronic spin coherence time approaching one second. *Nature Communications* **4**, 1743 (2013).

33. Koehl, W. F., Buckley, B. B., Heremans, F. J., Calusine, G. & Awschalom, D. D. Room temperature coherent control of defect spin qubits in silicon carbide. *Nature* **479**, 84–7 (2011).

34. Falk, A. L. *et al.* Electrically and Mechanically Tunable Electron Spins in Silicon Carbide Color Centers. *Phys. Rev. Lett.* **112,** 187601 (2014).

35. Yu, P.-L., Purdy, T. P. & Regal, C. A. Control of Material Damping in High-Q Membrane Microresonators. *Phys. Rev. Lett.* **108,** 083603 (2012).

36. Jarmola, A., Acosta, V. M., Jensen, K., Chemerisov, S. & Budker, D. Temperature- and Magnetic-Field-Dependent Longitudinal Spin Relaxation in Nitrogen-Vacancy Ensembles in Diamond. *Phys. Rev. Lett.* **108**, 197601 (2012)

37. Rosskopf, T. *et al.* Investigation of Surface Magnetic Noise by Shallow Spins in Diamond. *Phys. Rev. Lett.* **112,** 147602 (2014).

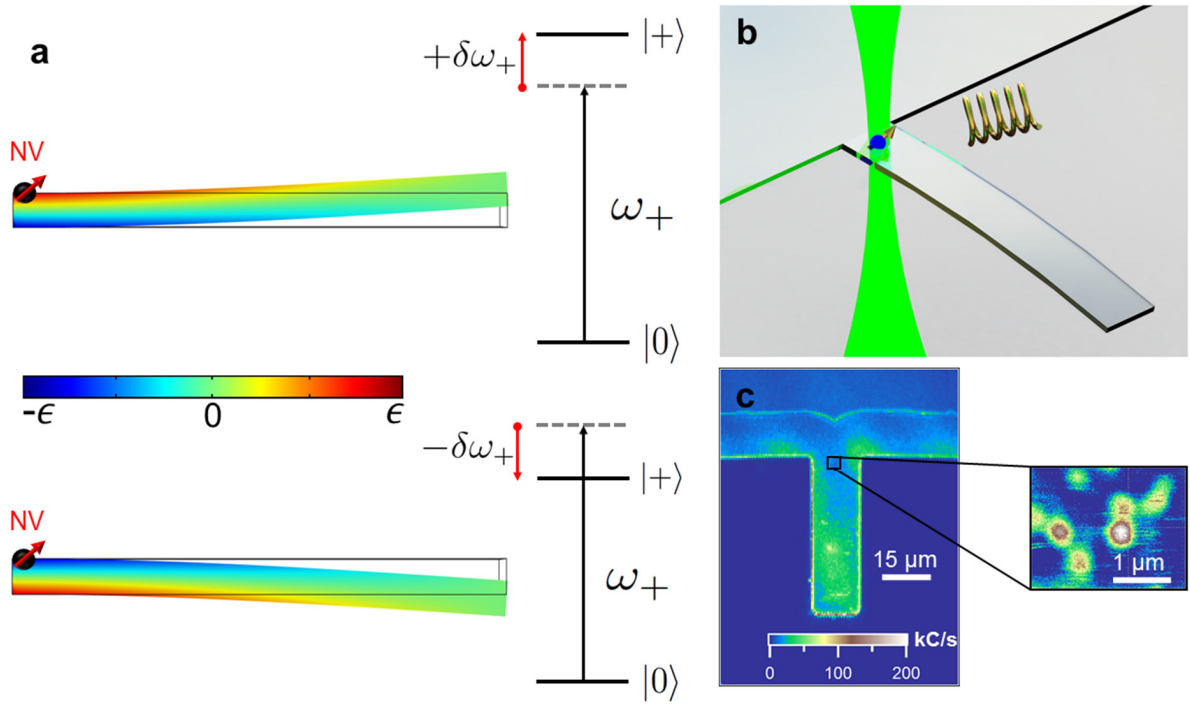

**Figure 1 A hybrid NV-cantilever system. a,** Strain modulates the spin levels of an NV center embedded in a cantilever. The strain profile was simulated using a finite element method. **b,** Schematic of experiment: a green laser is focused onto a single NV spin embedded in a cantilever via a confocal microscope for initialization and readout. Microwaves are used for pulsed spin manipulation. **c,** Confocal image of a cantilever showing the presence of single embedded NV centers.

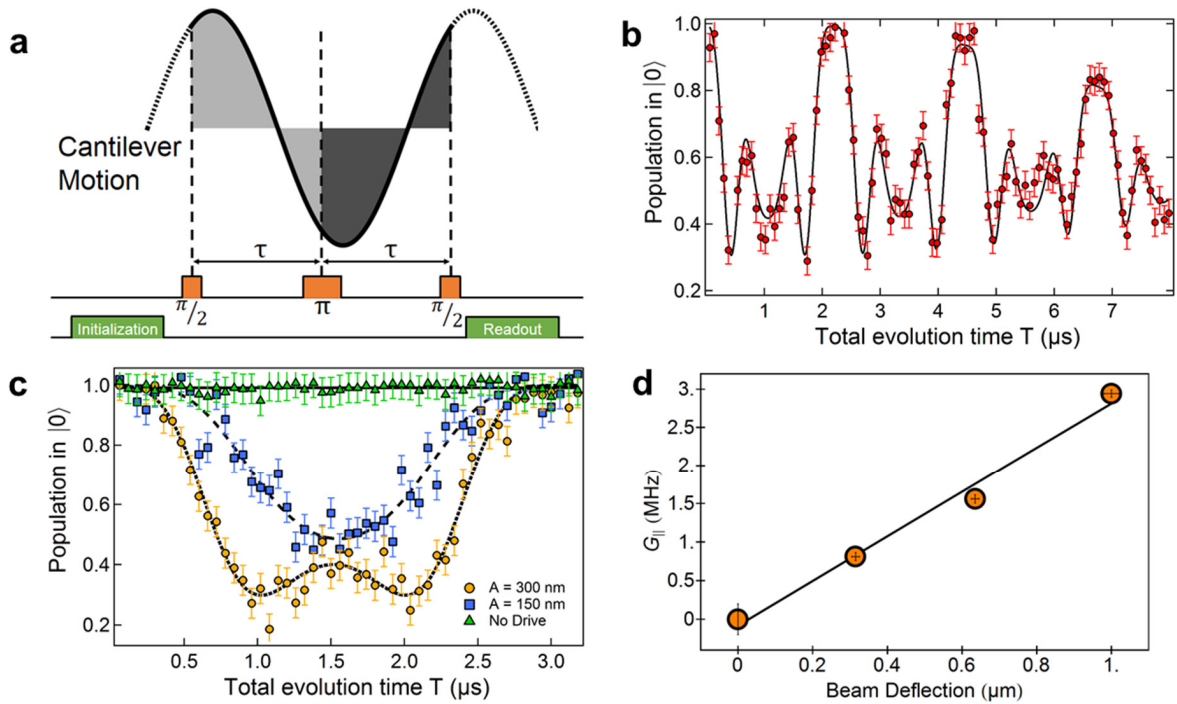

**Figure 2 Axial strain detection with a single NV. a,** AC axial strain is detected with a Hahn Echo sequence. Transverse strain is suppressed by an applied 22 G axial magnetic field. The $\pi$ pulse allows the phases from the first (light gray) and second (dark grey) free evolution periods to add constructively. **b,** Measured Hahn echo spin population (red circles) for an NV at the base of a cantilever with $\omega_m/2\pi = 884 \, kHz$ and an oscillation amplitude of 650 nm as a function of total evolution time $T = 2\tau$. The echo signal is given by a zero order Bessel function (fit is black solid line) because the cantilever motion is not phase-locked to the timing of the control pulses. Error bars correspond to standard error in photon counting. **c,** The amplitude of cantilever motion impacts the spin evolution. Hahn echo spin population of an NV at the base of a $\omega_m/2\pi = 630 \, kHz$ cantilever in the absence of drive (green triangles), with a drive corresponding to 150 nm cantilever amplitude (blue squares) and 300 nm amplitude (yellow circles). Dotted black lines are fits to the data. **d,** Plotted strain coupling $G_\parallel$ for an NV 11 μm from the base on the cantilever described in (b) as a function of beam deflection. Vertical error bars correspond to standard error of the fit to the theoretical echo signal. Horizontal error bars correspond to the uncertainty in beam deflection from interferometric measurements.

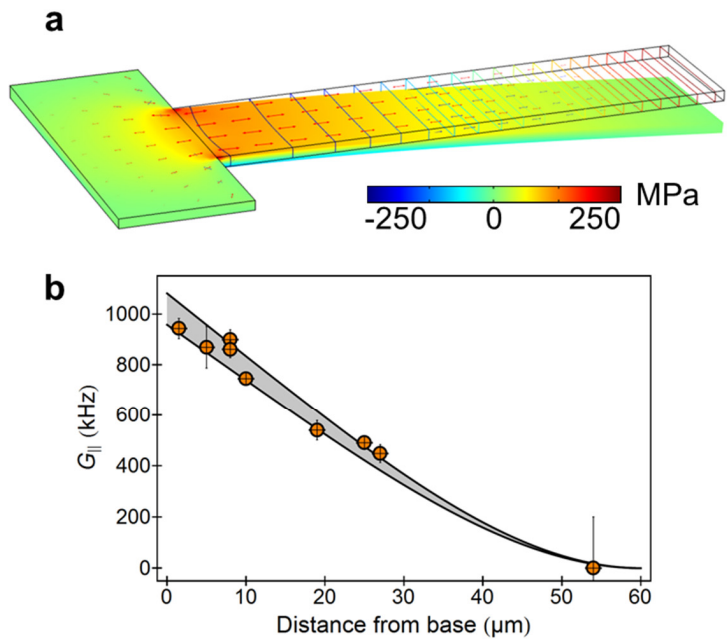

**Figure 3 Single-spin strain imaging. a,** Stress profile for cantilever used in this experiment at 250 nm of beam deflection using a finite element method simulation. **b,** Measured strain coupling (orange circles) as a function of the NV's distance from the cantilever base for a fixed 250 nm oscillation amplitude. The grey shaded area shows the region of expected strain couplings from theory including uncertainties in NV depth (13 nm) and amplitude of driven motion (10 nm). Vertical error bars correspond to the standard error from fits to the expected control sequence signal. Horizontal error bars correspond to the uncertainty in the NV's lateral position.

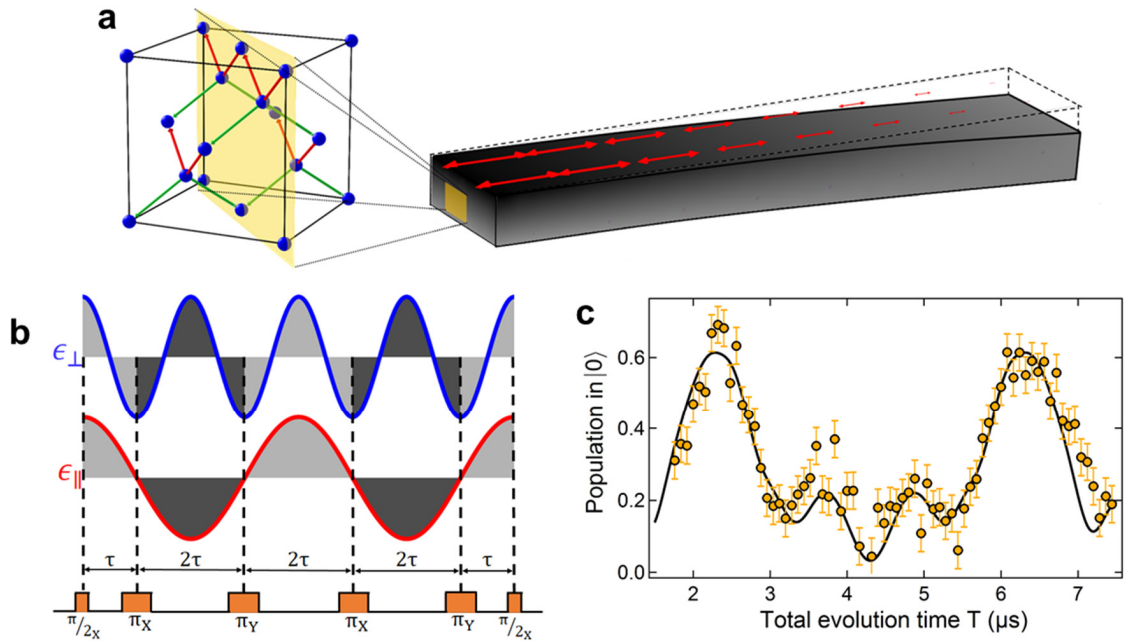

**Figure 4 Measurement of transverse strain. a,** Stress profile for a cantilever during bending (red arrows). The face of the cantilever (gold square) corresponds to the (110) plane. NVs oriented along [$\bar{1}1\bar{1}$] and [$1\bar{1}\bar{1}$] (red bonds) experience mostly transverse strain, with a small axial strain due to the Poisson effect[23]. NVs oriented along [111] and [$\bar{1}\bar{1}1$] (green bonds) experience predominantly axial strain (see Supplementary Information). Measurements of transverse strain are done with NVs oriented [$\bar{1}1\bar{1}$] and [$1\bar{1}\bar{1}$]. **b,** XY-4 pulse sequence used to measure transverse strain. Transverse strain (in blue) modulates the qubit splitting approximately twice as fast as axial strain (in red). **c,** XY-4 spin population (see Supplementary Information) for an NV (yellow circles) 3 μm from the base with a beam deflection of 675 nm. The expected XY-4 signal for a model considering both axial and transverse strain is shown in black. Error bars correspond to photon counting noise.

# Supplementary Information

**Dynamic strain-mediated coupling of a single diamond spin to a mechanical resonator**

P. Ovartchaiyapong†, K.W. Lee†, B.A. Myers, A.C. Bleszynski Jayich*

†These authors contributed equally to this work

*To whom correspondence should be addressed. Email: ania@physics.ucsb.edu

## Supplementary Note 1. The NV-strain interaction

Strain-induced displacements of atoms in the diamond crystal alter the electron density and generate local electric fields[1]. In this model of strain, the Hamiltonian obeys ($h = 1$)

$$H_{NV} = (D_0 + d_\parallel \epsilon_z)S_z^2 + \gamma_{NV} \mathbf{S} \cdot \mathbf{B} - d_\perp [\epsilon_x(S_x S_y + S_y S_x) + \epsilon_y(S_x^2 - S_y^2)] \quad (1)$$

Here, $D_0$ is the zero field splitting ($D_0 \approx 2.87$ GHz), $\gamma_{NV}$ is the gyromagnetic ratio ($\approx$ 2.8 MHz $\cdot$ G$^{-1}$), $\mathbf{B}$ is the applied DC magnetic field, $\{S_i\}_{i=x,y,z}$ are the spin 1 representation of the Pauli matrices, $\{\epsilon_i\}_{i=x,y,z}$ are the diagonal components of the strain tensor in the NV's coordinate basis, and $d_\parallel$ and $d_\perp$ are defined as the strain susceptibility parameters parallel and perpendicular to the NV symmetry axis. The strain induced energy shift has been measured in the excited state of the NV center[2] ($d_{es} = 1.2$ PHz $\cdot$ strain$^{-1}$), but the ground state strain susceptibilities have yet to be characterized.

Because stress and strain are tensors that are related by an elasticity matrix, a uniaxial stress along one particular direction will also induce strains transverse to the direction of applied stress due to the Poisson effect. Including this effect in our analysis is crucial for characterizing the strain interaction. Our model does not take into account shear strain, but our measurements indicate that shear does not couple significantly to the ground state spin levels of the NV center. However, we note that the cantilever-based technique presented here can be extended in future experiments to probe effects of shear on the NV. As we will show in a later section, the shear experienced by all four NV orientations can be calculated from geometric considerations. In general, shear will break all symmetries of the NV center, and the shear-spin interaction may enable interesting quantum applications in the future.

To qualitatively understand the spin-strain interaction, it is useful to see how uniaxial strains affect the $C_{3v}$ symmetry of the NV. A uniaxial strain along the NV's symmetry axis, $\epsilon_z$, results in a dilation or contraction of the nitrogen bond, which transforms as the $A_1$ irreducible representation of the $C_{3v}$ group, preserving all symmetries. Therefore, we would not expect any breaking of the $m_s = \pm 1$ degeneracy, and moreover, we expect that the axial strain should commute with the zero field NV Hamiltonian. Uniaxial strains in the plane perpendicular to the symmetry axis of the NV dilate and contract carbon bonds which break the $C_3$ rotational symmetry and the $\sigma_v$ reflection symmetry. Thus, transverse strains transform as the $E$ irreducible representation of the $C_{3v}$ group. Therefore, we expect the degeneracy of the $m_s = \pm 1$ levels to break and moreover, that these spin levels will mix. The Hamiltonian may be written to more directly reflect these qualitative changes by using the spin raising and lowering operators and defining the axial strain, $\epsilon_\parallel = \epsilon_z$, and transverse strain $\epsilon_\perp = \sqrt{\epsilon_x^2 + \epsilon_y^2}$. The Hamiltonian simplifies to

$$H_{\text{NV}} = (D_0 + d_\parallel \epsilon_\parallel) S_z^2 + \gamma_{\text{NV}} \boldsymbol{S} \cdot \boldsymbol{B} - \frac{d_\perp \epsilon_\perp}{2}\left(e^{-i\phi_s} S_+^2 + e^{i\phi_s} S_-^2\right) \tag{2}$$

where $\tan\phi_s = \epsilon_y/\epsilon_x$. We see while axial strain shifts the $|\pm 1\rangle$ states together relative to the $|0\rangle$ state, transverse strain mixes the $|\pm 1\rangle$ levels. In our experiment, NVs embedded in a SCD cantilever experience a strain when the cantilever bends, and for small displacements of the cantilever, such as in this work, the strain is linear with displacement.

We define the strain due to the fundamental mode to be $\epsilon = \epsilon_0 \frac{X_d}{x_0}$, where $X_d$ is the amplitude of driven motion and $x_0$ is the amplitude of zero point motion for the fundamental mode of the resonator, and $\epsilon_0$ is the strain induced by the zero point motion. Going forward, we define $g_\parallel = d_\parallel \epsilon_{0\parallel}$ and $g_\perp = d_\perp \epsilon_{0\perp}$, $X = X_d/x_0$, $G_\parallel = g_\parallel X$ and $G_\perp = g_\perp X$.

In our experiments, we introduce a static magnetic field that is closely aligned to the NV axis in order to split the degeneracy of the $|\pm 1\rangle$ spin levels. In the low perpendicular field limit, $(\gamma_{NV}B_\perp)^2/D_0 \ll \sqrt{(\gamma_{NV}B_z)^2 + (d_\perp\epsilon_\perp)^2}$, and the spin eigenstates are approximately given by

$$|0\rangle = |0\rangle \tag{3}$$

$$|+\rangle = \cos\frac{\alpha}{2}|1\rangle - e^{-i\phi_s}\sin\frac{\alpha}{2}|-1\rangle \tag{4}$$

$$|-\rangle = \sin\frac{\alpha}{2}|1\rangle + e^{-i\phi_s}\cos\frac{\alpha}{2}|-1\rangle \tag{5}$$

where $\alpha/2$ is the Stückelberg angle and $\tan\alpha = d_\perp\epsilon_\perp/\gamma_{NV}B_z$. Note that in the absence of transverse strain this returns the usual $S_z$ eigenstates.

The corresponding eigenenergies are

$$E_0 \simeq 0$$

$$E_\pm(X) \simeq D_0 + g_\parallel X \pm \sqrt{(\gamma_{NV}B_z)^2 + (g_\perp X)^2} \tag{6}$$

In our experiment, we encode our spin qubit in the $\{|0\rangle, |+\rangle\}$ spin states with a qubit frequency of $\nu_+(X) \simeq D_0 + g_\parallel X + \sqrt{(\gamma_{NV}B_z)^2 + (g_\perp X)^2}$. This expression shows how we can distinguish the effects of axial and transverse strain; an AC axial strain generated by the cantilever modulates the qubit splitting at the cantilever frequency whereas an AC transverse strain modulates at nearly twice the frequency.

**Supplementary Note 2. Hahn echo and axial strain signal**

Here, we derive the theoretical Hahn Echo signal for an NV under AC strain. We treat the cantilever as a quantum harmonic oscillator. Because the cantilever is driven on resonance, we may neglect any dissipation in the resonator. This treatment returns the classical result in the limit that $\hbar\omega_m \ll k_b T$. We assume the NV is in the presence of a large external magnetic field aligned to the symmetry axis. The $\{|0\rangle, |+\rangle\}$ states are selectively addressed by applying near-resonant microwaves at the $|0\rangle \leftrightarrow |+\rangle$ transition frequency to the NV, allowing the $|-\rangle$ state to be ignored due to the large microwave detuning.

An axial strain in the $\{|0\rangle, |+\rangle\}$ basis coherently modulates the system energy at the cantilever frequency. If $\left(\frac{G_\perp}{2\gamma_{NV}B_z}\right)^2 \ll 1$, then the transverse strain will have a negligible effect on the spin evolution in the experiment. Therefore, we model the system with the spin-boson Hamiltonian ($\hbar = 1$),

$$H = \omega_+ \sigma_z + \omega_m a^\dagger a + 2\pi g_\parallel (a + a^\dagger)\sigma_z \tag{17}$$

where $\omega_+ = 2\pi(D_0 + \gamma_{NV}B_z)$ is the transition frequency in the absence of strain, $\omega_m$ is the cantilever frequency and $a$ is the annihilation operator of the oscillator. To readout the spin, a green laser is applied to the NV and photons emitted into the phonon sideband are collected. The readout sequence is calibrated by a repolarization pulse that measures the fluorescence of the $|0\rangle$ state. Thus, we are interested in the probability that the NV's final state is $|0\rangle$, which is given by $P(2\tau) = \frac{1}{2}(1 + \cos(\theta(2\tau)))$, where $\theta$ refers to the phase accumulated in the echo sequence and $2\tau$ is the total free evolution time of the spin.

In generic dynamical decoupling pulse sequences such as a Hahn echo, a $\pi$ pulse inserted into the free evolution of the spin acts to swap $\sigma_z$ to $-\sigma_z$ and vice versa. In the spin's reference frame, the sign of the interaction then "toggles" each time a $\pi$ pulse is applied. Therefore, it is convenient to write out the Hamiltonian in the interaction picture with respect to $\omega_+ \sigma_z + \omega_m a^\dagger a$ and introduce the "toggling frame" with

$$H_I = 2\pi g_\parallel X(t)\sigma_z f(t,\tau) \tag{18}$$

where $X(t) = e^{i\omega_m a^\dagger a t}(a + a^\dagger)e^{-i\omega_m a^\dagger a t}$ and $f(t,\tau)$ is a square wave that switches between $\pm 1$ and defines the toggling frame generated by the $\pi$ pulse and $\tau$ defines the length of a free evolution period. Because we have neglected dissipation, $X(t) = X\cos(\omega_m t + \phi)$. The phase acquired for a Hahn Echo experiment with total evolution time $T = 2\tau$ is given by

$$\theta(T = 2\tau) = \int_t^{t+2\tau} dt' 2\pi G_\parallel \cos(\omega_m t' + \phi) f(t', \tau) \qquad (19)$$

The timing of each Hahn echo experiment is not phase locked with the cantilever's motion, thus the signal $P(T)$ is averaged of a uniformly distributed initial phase of the cantilever, $\phi$. For the overall probability to be in the $|0\rangle$ state, we obtain

$$P(T) = \frac{1}{2}\left[1 + J_0\left(\frac{8\pi G_\parallel}{\omega_m}\sin^2\left(\frac{\omega_m T}{4}\right)\right)\right] \qquad (20)$$

where $J_0(x)$ is the $0^{th}$ order Bessel function of the first kind.

If we take into account the $T_2$ decay of the NV, the probability to be in the $|0\rangle$ state obeys

$$P(T) = \frac{1}{2}\left[1 + e^{-(T/T_2)^n} J_0\left(\frac{8\pi G_\parallel}{\omega_m}\sin^2\left(\frac{\omega_m T}{4}\right)\right)\right] \qquad (21)$$

We note that several values of $n \in [1,3]$ yield good fits to the data, but the extracted decay constant, $T_2$, is relatively unchanged for any of these exponents.

This probability is then mapped to the normalized fluorescence and hence the populations, giving the overall expected signal. We use this expression to extract the axial strain coupling parameter, $G_\parallel$.

## Supplementary Note 3. Strain in a singly-clamped cantilever

Here, we derive the strain experienced by an NV due to the bending of the cantilever. This can be done with elasticity theory and Euler-Bernoulli flexure theory, which is used to describe the bending modes of a long, thin beam. Consider a singly-clamped diamond beam of length $l$, width $w$, and thickness $t$. The wave equation for beam deflections of the neutral axis obeys

$$EI\frac{\partial^4 U}{\partial z^4} = -\rho A \frac{\partial^2 U}{\partial t^2} \qquad (7)$$

where $U$ represents the beam deflection in the $y$ direction and $z$ is along the length of the cantilever, as shown in Supplementary Fig. 1a. Here, E is Young's modulus of diamond ($\approx 1.2$ TPa), $I = \frac{wt^3}{12}$ is the moment of inertia of the beam, $\rho$ is the mass density of diamond (3.5 g · cm$^{-3}$), and $A = tw$ is the cross sectional area in the transverse plane. The solutions of (S7) for a singly-clamped beam take the form $U_n(t,z) = u_n(z)e^{-i\omega_n t}$, where $u_n(z)$ is given by

$$u_n(z) = a_n(\cos(\beta_n z) - \cosh(\beta_n z)) \qquad (8)$$
$$+ b_n(\sin(\beta_n z) - \sinh(\beta_n z))$$

which obeys the boundary conditions $u(0) = u'(0) = u''(l) = u'''(l) = 0$. The coefficients $a_n$ and $b_n$ satisfy the condition $a_n/b_n = -1.3622, -0.09819, -1.008, -1.000, ...$ and $\beta_n$ satisfies the condition $\cos(\beta_n l)\cosh(\beta_n l) = -1$.

The eigenfrequencies of the cantilever are given by

$$\omega_n = \beta_n^2 \sqrt{\frac{EI}{\rho A}} = (\beta_n l)^2 \sqrt{\frac{E}{12\rho} \frac{t}{l^2}} \qquad (9)$$

In this paper, we are interested in the fundamental mode of the cantilever where $n = 1$ and $\beta_1 l = 1.875$. To normalize the fundamental mode solution, we set the value of $u_1(l)$ to be equal to the amplitude of zero point motion, $x_0 = \sqrt{\frac{\hbar}{2\mu\omega_1}}$, where $\mu = \frac{1}{4}ltw\rho$ is the effective mass of the cantilever

If the NV sits at a point $R_0$ from the neutral axis of the beam, which here is defined as the line passing through the center of the beam parallel to $z$, then the strain associated with the fundamental mode of the cantilever as a function of $z$ is given by $\epsilon(z) \simeq -R_0 \frac{\partial^2 u_1}{\partial z^2}$, thus we find

$$\epsilon(z) \simeq R_0 x_0 \frac{1.75}{l^2}[\cos\left(1.875\frac{z}{l}\right) + \cosh\left(1.875\frac{z}{l}\right)$$
$$- \frac{1}{1.3622}\left(\sin\left(1.875\frac{z}{l}\right) + \sinh\left(1.875\frac{z}{l}\right)\right)] \qquad (10)$$

In our cantilevers, $R_0 = t/2 - d_i$, where $d_i$ is the nitrogen implantation depth of approximately 50 nm. Note that the strain is maximized at the base of the cantilever and decreases along the length of the cantilever, where it reaches a value of 0 at the tip.

**Supplementary Note 4. Calculation of the strain tensors**

To extract the strain coupling parameters in Supplementary Equation (2), it is necessary to transform the strain tensor from the cantilever's basis to the NV's basis. Here, we define the $z$ axis to be the cantilever axis, the $y$ axis to be perpendicular to the plane of the cantilever and the $x$ axis to be in the plane of the cantilever, as depicted in Supplementary Fig. 1a. We begin with the strain tensor defined in the cantilever's basis, with $\epsilon(z)$ defining the unaxial strain along the cantilever's axis due to the bending defined in Supplementary Equation (10). The strain tensor in the cantilever's basis is then

$$\Xi = \begin{bmatrix} -\nu\epsilon(z) & 0 & 0 \\ 0 & -\nu\epsilon(z) & 0 \\ 0 & 0 & \epsilon(z) \end{bmatrix} \quad (11)$$

where $\nu = .069$ is the Poisson ratio for CVD grown diamond[3].

The cantilevers are cut along the (110) plane and the diamond is grown along the (100) plane. Therefore, the $z$ axis is parallel to the [110] direction, the $y$ axis is parallel to the [100] direction, and the $x$ axis is parallel to the [1$\bar{1}$0]. To obtain the strain tensor for the 4 different orientations, we apply a transformation of the form $\mathbf{Q}\Xi\mathbf{Q}^\mathbf{T}$ that takes the principal axes of the cantilever into the principal axes of the NV center. Specifically, the $z$ axis of the cantilever will be transformed into the NV symmetry axis and the $x$ axis of the cantilever will be transformed so that it lies along the projection of a carbon bond in the plane perpendicular to the NV axis, as shown in Supplementary Fig. 1b. We note that because of the $C_3$ rotational symmetry, there will be 3 unique transformations for each orientation. However, the overall transverse and axial strains will remain invariant, and only the shear terms and the strain phase $\phi_s$ will be affected.

We neglect shear in our analysis and $\phi_s$ will not affect the energy or magnitude of strain.

NVs with their symmetry axes along the $[\bar{1}1\bar{1}]$ and $[1\bar{1}\bar{1}]$ directions will experience the same strain, and NVs with their symmetry axes along the $[111]$ and $[\bar{1}\bar{1}1]$ directions will experience the same strain. The transformations for the NVs oriented along the $[111]$ and $[\bar{1}\bar{1}1]$ directions are $\mathbf{Q}_{111} = \mathbf{R}_z(180°)\mathbf{R}_y(54.5°)\mathbf{R}_x(90°)$ and $\mathbf{Q}_{\bar{1}\bar{1}1} = \mathbf{R}_z(0°)\mathbf{R}_y(-54.5°)\mathbf{R}_x(90°)$, where $\mathbf{R}_i(\theta)$ is the SO(3) rotation matrix about the $i$th cantilever axis. The corresponding transformation for NVs oriented along the $[\bar{1}1\bar{1}]$ and $[1\bar{1}\bar{1}]$ directions are $\mathbf{Q}_{\bar{1}1\bar{1}} = \mathbf{R}_z(-90°)\mathbf{R}_x(35°)$ and $\mathbf{Q}_{1\bar{1}\bar{1}} = \mathbf{R}_z(90°)\mathbf{R}_x(-35°)$. We obtain the following transformed strain tensors

$$\Xi_{[\bar{1}1\bar{1}],[1\bar{1}\bar{1}]} = \begin{bmatrix} \epsilon(z) & 0 & 0 \\ 0 & -\nu\epsilon(z) & 0 \\ 0 & 0 & -\nu\epsilon(z) \end{bmatrix} \tag{12}$$

$$\Xi_{[111]} = \begin{bmatrix} (.33-.67\nu)\epsilon(z) & 0 & -.47(1+\nu)\epsilon(z) \\ 0 & -\nu\epsilon(z) & 0 \\ -.47(1+\nu)\epsilon(z) & 0 & (.67-.33\nu)\epsilon(z) \end{bmatrix} \tag{13}$$

$$\Xi_{[\bar{1}\bar{1}1]} = \begin{bmatrix} (.33-.67\nu)\epsilon(z) & 0 & .47(1+\nu)\epsilon(z) \\ 0 & -\nu\epsilon(z) & 0 \\ .47(1+\nu)\epsilon(z) & 0 & (.67-.33\nu)\epsilon(z) \end{bmatrix} \tag{14}$$

Using these strain tensors, we define the strain coupling parameters, $G_\parallel$ and $G_\perp$ to be

$$G_\parallel(z) = d_\parallel \frac{X_d}{x_0} \Xi_{zz}(z) \tag{15}$$

$$G_\perp(z) = d_\perp \frac{X_d}{x_0} \sqrt{(\Xi_{xx}(z))^2 + (\Xi_{yy}(z))^2} \tag{16}$$

where $X_d$ is the amplitude of driven motion.

To extract the strain susceptibility parameters, we first fit our pulsed magnetic resonance data to the theoretically predicted coherence envelopes, which gives a value for each coupling $G_\parallel$ and $G_\perp$ with an uncertainty that is given by the standard error in the fit. Next, we fit these values as a function of $z$ to the expression given by (S10) including uncertainties in the implantation depth and drive amplitude. From this fit, we extract the strain susceptibilities with an uncertainty given by the standard error in the fit. Finally, our quoted value of the axial strain susceptibility is given by averaging the extracted values of the susceptibilites for each measured NV, with the quoted uncertainty given by the standard error of the mean. The value and uncertainty of the transverse strain susceptibility is given by a measurement of a single NV.

In Fig. 3b of the main text, the shaded region is given by Supplementary Equation (15) for the axial strain including uncertainties in the NV's depth and the amplitude of driven motion.

**Supplementary Note 5. Strain sensitivity estimation**

Here, we provide a shot-noise limited estimation of the strain sensitivity of our device following a similar approach provided in Supplementary References 4 and 5. We define the strain sensitivity $\eta_s$ as the sensitivity of a strain measurement with a signal to noise ratio of one. The strain measurement to be considered is a Hahn echo experiment in the $\{|0\rangle, |+\rangle\}$ basis in which an AC strain is introduced from sinusoidal motion of the resonator. The free evolution period is matched to half of the resonator's oscillation period and phase locked to the motion, maximizing the accumulated phase. In our strain sensing measurements and in the analysis that follows, we focus solely on the NV sensitivity to axial strain, but note that this analysis may also be applied for transverse strain.

The strain sensitivity may be written as

$$\eta_s = \frac{2\pi A}{d_\parallel t} e^{(t/T_2)^n} \quad (22)$$

where $A$ is defined as

$$A = \sqrt{\frac{2(1+C)}{N\alpha_0(1-C)^2}} \tag{23}$$

$t$ is the total Hahn echo measurement time, $N$ is the total number of measurements, $n$ is the exponent given in Supplementary Equation (21) which is determined by the decoherence mechanism for the spin, $T_2$ is the Hahn echo spin coherence time, $\alpha_0$ is the number of photons detected by the $|0\rangle$ state in a single Hahn echo shot, and $C$ gives the fluorescence contrast between the $|0\rangle$ and $|+\rangle$ states. The value of $\alpha_0$ is primarily determined by the collection efficiency of the confocal microscope. In the experiment here, the readout sequence is 350 ns long, leading to an $\alpha_0$ of .01 photons. Furthermore, the fluorescence contrast yields a value of $C = 0.85$. We note that higher collection effiencies and contrasts have been achieved with high numerical aperture microscope objectives. A typical Hahn echo $T_2$ coherence time for NVs in our current devices is 10 μs. In the future, the $T_2$ coherence times of NVs in these structures may be significantly increased with nitrogen delta-doping techniques, which consistently produce shallow NVs with excellent coherence properties (see ref. 30 from the main text). As mentioned previously, several values of $n$ in the range [1,3] yielded good fits to our data, and henceforth we will set $n$ equal to one.

Assuming no dead time between measurements, $N = T/t$, where $T$ is the total interrogation time. In addition, it can be shown [S5] that the signal to noise ratio is maximized when $t = T_2/2$, and hence the sensitivity and minimum detectable strain can be written as

$$\eta_s = \frac{2\pi\sqrt{2}A}{d_\parallel \sqrt{TT_2}} e^{1/2} \tag{24}$$

Substituting the measured values of all parameters from our devices, we extract a minimum detectable strain of $3 \times 10^{-6}$ in a 1 second measurement, implying the sensitivity of our device is $3 \times 10^{-6}$ strain $\cdot$ Hz$^{-1/2}$. Given the minimum detectable strain, the minimum detectable amplitude of motion may be then inferred using Supplementary Equation (15), from which we estimate a minimum detectable amplitude of motion of 7 nm.

**Supplementary Note 6. XY-4 and transverse strain signal**

To further enhance the sensitivity to strain, we use the higher order dynamical decoupling pulse sequence, XY-4. By introducing multiple $\pi$ pulses, the spin may acquire more phase and simultaneously decouple from higher frequency sources of decoherence. The XY-4 pulse obeys the timing described for the canonical CPMG pulse sequences. However, the XY-4 control propagator alternates the rotation axes of the $\pi$ pulses between $x$ and $y$, which serves two primary functions. First, this corrects small timing errors in the microwave pulses to first order. Second and more importantly, this facilitates delibrate phase accumulation, which will be the foundation of our signal. The timing for the XY-4 sequence used in the main text is given by $\frac{\pi}{2}_x - \tau - \pi_x - 2\tau - \pi_y - 2\tau - \pi_x - 2\tau - \pi_y - \tau - \frac{\pi}{2}_x$, where the total evolution time is $T = 8\tau$.

We use the XY-4 sequence specifically for measurements of the axial strain coupling for NVs farther down the cantilever, which experience relatively small strain. We also use this sequence to measure the transverse strain coupling. In the $\{|0\rangle, |+\rangle\}$ basis, transverse strain is essentially a second order correction to the energy and is suppressed by the axial magnetic field, making it difficult to detect with the usual Hahn Echo sequence. We note, however, that this enhancement of the transverse strain only has a measurable effect on the spin evolution for the class of NVs that experience predominantly transverse strain.

In our measurements of the axial strain coupling, we only need to consider the axial strain interaction. The Hamiltonian in the interaction picture is again given by Supplementary Equation (18), where the toggling frame defined by $f(t,\tau)$ now reflects the timing of the XY-4 sequence. Using the method described in the previous section, we find that the probability to be in the $|0\rangle$ state is given by

$$P(T) = \frac{1}{2}[1 - e^{-(T/T_2)^n} J_0(\frac{4\pi G_\parallel}{\omega_m}\sin\left(\frac{\omega_m T}{2}\right) \times (1 - \sec(\frac{\omega_m T}{8})))] \qquad (25)$$

A typical XY-4 signal for an NV that experiences predominantly axial strain is shown in Supplementary Fig. 2. This NV was located at the base of a cantilever that was being driven with an oscillation amplitude of 250 nm. From the fit shown in Supplementary Fig. 2, we extract an axial strain coupling $G_\parallel = 1.1$ MHz and a cantilever frequency of $889 \pm 2$ kHz, in agreement with the driving frequency of 884.894 kHz.

For NVs in the high transverse strain class, the Hamiltonian in the interaction picture is given by

$$H_I = 2\pi[(G_\parallel\cos(\omega_m t + \phi) + \sqrt{(\gamma_{NV}B_z)^2 + G_\perp^2\cos^2(\omega_m t + \phi)}\,]f(t,\tau)\sigma_z \qquad (26)$$

The signal will be given by $P(T) = \frac{1}{2}\left[1 - e^{-(T/T_2)^n}\cos(\theta_\parallel(T) + \theta_\perp(T))\right]$, where $\theta_\parallel = \frac{2G_\parallel}{\omega_m}\sin(\frac{\omega_m T}{2})(1 - \sec(\frac{\omega_m T}{8}))\cos(\omega_m T + \phi)$. To evaluate $\theta_\perp$ we use the incomplete elliptic integral of the second kind $E(m,k)$. If the elliptic modulus, $k$, satisifies $0 < k^2 < 1$, then we define the incomplete elliptic integral of the second kind to be $E(m,k) = \int_0^m dx\sqrt{1 - k^2\sin^2 x}$. The perpendicular phase is obtained to be

$$\theta_\perp = \frac{2\pi R}{\omega_m}[2E\left(\frac{\omega_m T}{8} + \phi, \frac{G_\perp}{R}\right) - E\left(\phi, \frac{G_\perp}{R}\right)$$

$$-2E\left(\frac{3\omega_m T}{8} + \phi, \frac{G_\perp}{R}\right) + 2E\left(\frac{5\omega_m T}{8} + \phi, \frac{G_\perp}{R}\right)$$

$$-2E\left(\frac{7\omega_m T}{8} + \phi, \frac{G_\perp}{R}\right) + E\left(\omega_m T + \phi, \frac{G_\perp}{R}\right)] \qquad (27)$$

where we have defined $R = \sqrt{(\gamma_{NV}B_z)^2 + G_\perp^2}$ for convenience.

Averaging over the initial phase of the cantilever, we obtain the XY-4 signal. Using this expression, we extract the transverse strain coupling. We note that this expression is not analytic. Therefore, to extract $G_\perp$ we used a series of numerically integrated fit functions that sample various values of $\omega_m$, $g_\parallel$ and $g_\perp$ while keeping the Zeeman splitting and amplitude of driven motion constant. Using least-squares fitting, we are able to extract the values of $\omega_m$, $g_\parallel$ and $g_\perp$ that give the best fit. As a consistency check, the extracted value of $g_\parallel$ can be compared to the expected value given the measured $d_\parallel$ and the expected strain given the strain tensors calculated in Supplementary Equations (12)-(14).

## Supplementary References

1. Van Oort, E. & Glasbeek, M. Electric-field-induced modulation of spin echoes of N-V centers in diamond. *Chem. Phys. Lett.* **168,** 529–532 (1990).

2. Togan, E. *et al.* Quantum entanglement between an optical photon and a solid-state spin qubit. *Nature* **466,** 730–4 (2010).

3. Klein, C. a. & Cardinale, G. F. Young's modulus and Poisson's ratio of CVD diamond. *Diam. Relat. Mater.* **2,** 918–923 (1993).

4. Falk, A. L. *et al.* Electrically and Mechanically Tunable Electron Spins in Silicon Carbide Color Centers. *Phys. Rev. Lett.* **112,** 187601 (2014).

5. Taylor, J. M. *et al.* High-sensitivity diamond magnetometer with nanoscale resolution. *Nat. Phys.* **4,** 810–816 (2008).

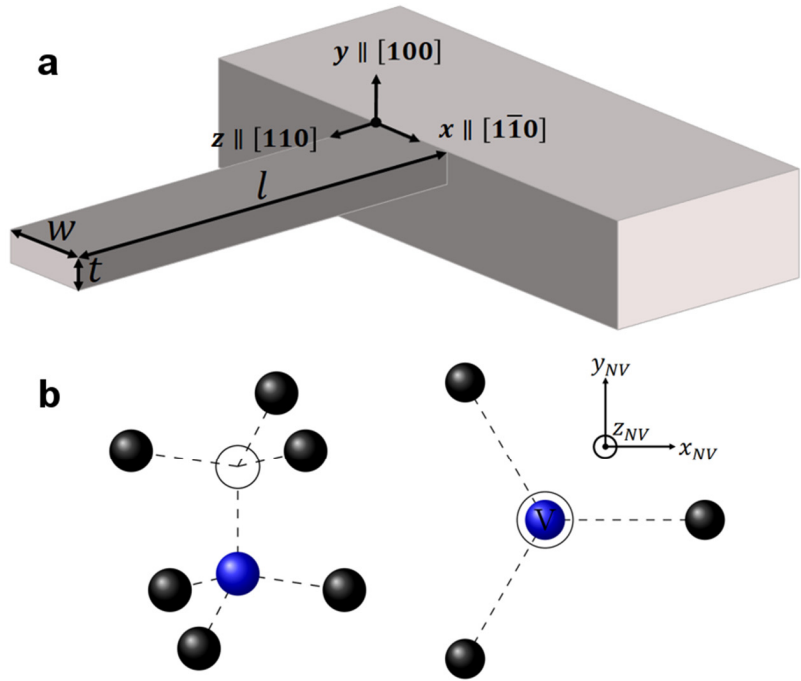

**Supplementary Figure 1 Principal axes of the cantilever and NV center. a,** Schematic of the cantilever. The principal axes of the cantilever are defined in terms of the diamond crystallographic directions. **b,** On the left is the NV structure. Black atoms represent carbon, the blue atom represents nitrogen and the white atom represents a vacancy. On the right, the NV is being viewed along its symmetry axis from the nitrogen to the vacancy.

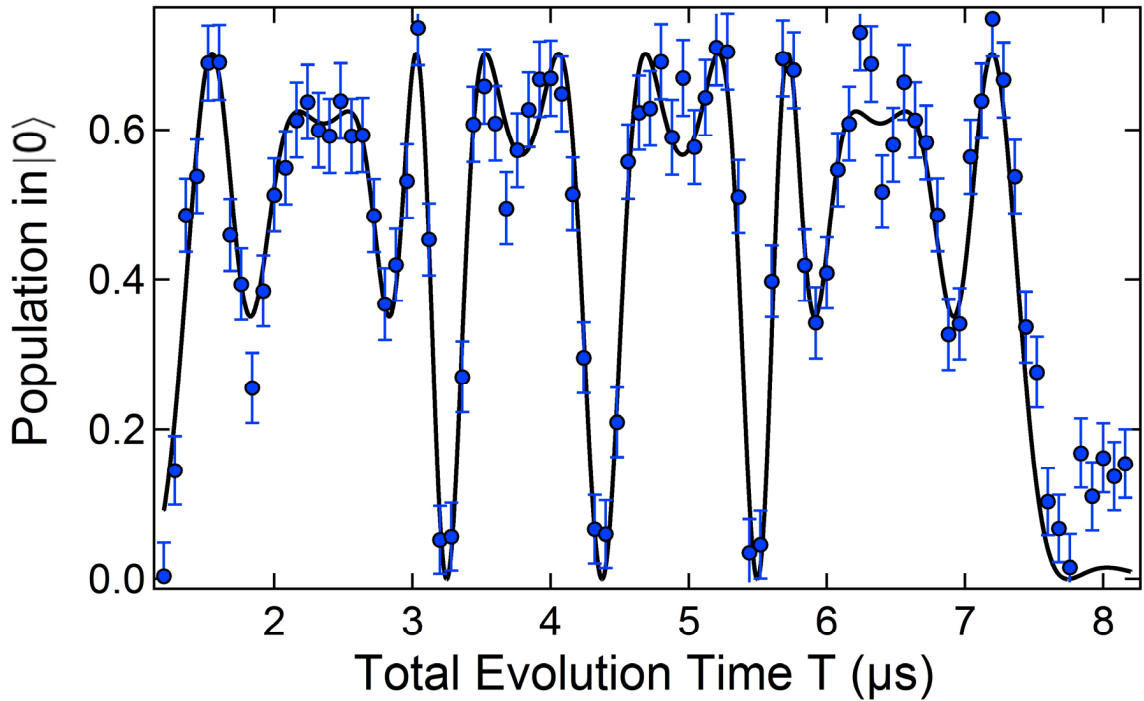

**Supplementary Figure 2 Axial strain detection using XY-4** Measured XY-4 spin population (blue circles) of an NV located at the base of a cantilever with $\omega_m/2\pi = 884.894$ kHz and an oscillation amplitude of 250 nm as a function of the total free evolution time. A fit to the expected XY-4 signal for a model considering only axial strain is shown by the black line. The vertical error bars are given by photon counting noise.